\documentstyle[twocolumn,prb,aps]{revtex}

\begin{document}

\title{ Observation of  domain wall resistivity in $\rm SrRuO_3$}
 \author{    L. Klein,$^1$ Y. Kats,$^1$ A. F. Marshall,$^2$ J. W. Reiner,$^3$ 
\\ T. H. Geballe,$^3$ M. R. Beasley,$^{2,3}$ and A. Kapitulnik$^3$ \\  
{\small $^1$Physics Department, Bar Ilan University, Ramat Gan 52900, Israel} \\
{\small $^2$Center for Materials Research, Stanford University,
 Stanford, California 94305} \\
{\small $^3$Edward L. Ginzton Laboratories, Stanford University,
 Stanford, California 94305} }
\date{\today}

\maketitle

\begin{abstract}
$\rm SrRuO_3$ is an itinerant ferromagnet with $T_c \sim 150 \rm \ K$. 
When $\rm SrRuO_3$ is cooled through $T_c$ in zero applied magnetic field,  a stripe
domain  structure appears whose orientation is uniquely determined 
by the large uniaxial magnetocrystalline anisotropy.   We find that the
ferromagnetic domain walls clearly enhance the resisitivity of $\rm SrRuO_3$ and that 
the enhancement has different temperature dependence 
for currents  parallel and perpendicular to the domain walls.  We
discuss possible interpretations of our results.
\end{abstract}

 Ferromagnetic domain walls (DW) are interfaces between
magnetic domains whose magnetization points in different directions.
For decades, there have been considerable theoretical\cite{the1,the2,the3,the4,the5,gor} and 
experimental\cite{the3,exp1,exp2,exp3} efforts to elucidate
the effect of DW on electrical resistivity. Early theoretical works 
predicted positive domain wall resistivity (DWR) due to 
reflection of electrons from the DW.\cite{the1,the2} More recently, 
a different
mechanism for  positive DWR has been proposed
(based on spin-dependent scattering due to defects) which 
pointed out  the similarity between the
magnetic structure 
 induced by DW and that present
in Giant Magnetoresistance (GMR) devices.\cite{the3,the5} 
On the other hand,  other models predicted
 negative DWR due to the  destruction of electron weak
localization induced by dephasing at the domain wall\cite{the4}; or
predicted that DWR can have either sign, depending on the difference between the
spin-dependent scattering lifetimes of the charge carriers, based on a semiclassical
treatment.\cite{gor} 
The experimental situation is also quite confusing. Reports on observed positive DWR in
materials like Co,\cite{the3} were followed by  studies which 
  indicated that the additional resistivity was actually a bulk effect.\cite{exp2} Moreover,
there are now strong indications that domain walls in iron (previously believed to enhance
resistivity) actually decrease resistivity.\cite{exp2,exp3}  

Contrary to most previous studies of DWR that focused on the elemental $3d$
ferromagnets,
we present  here a study of DWR in $\rm SrRuO_3$ 
which is a metallic perovskite and a 
$4d$ itinerant ferromagnet ($T_c \sim \rm 150\ K$). 
We find large  {\it positive} DWR (interface resistance of
$\sim 10^{-15} \ \Omega \rm \ m^2$ at low temperatures)
and present unique detailed temperature dependence
of DWR 
 for currents parallel
and perpendicular to the DW.  
 These results are important 
for better understanding the mechanisms involved in DWR
and since DW in this compound are extremely narrow\cite{loss} ($10 \rm \ \AA$),
 a justified comparison can be made
with models proposed for magnetic multilayers. 

$\rm SrRuO_3$ has special advantages for studying DWR for
its large uniaxial magnetocrystalline anisotropy field ($\sim 10 \rm \ T$) 
combined with its much smaller self
field ($4\pi M \sim \rm 0.2 \ T$). 
The large uniaxial anisotropy induces stripe magnetic structure (see inset in Figure 1) which
enables the study of DWR for currents parallel and perpendicular to the domain walls; 
furthermore, it is also responsible for the 
 narrow width of the DW which contributes to  the large magnitude of the
DWR. The combination of the large uniaxial anisotropy with a much smaller self field not only enables
the stability of a saturated magnetization state at zero applied field but it also
prevents the creation of closure domains near the sample surface when the sample is in its domain
state. As we discuss below, these two properties enable unequivocal determination
of DWR and avoid the 
need to consider bulk effects that may lead to mistaken or ambiguous identification of DWR. 

Our measurements were done on high-quality
single-crystal films  of $\rm SrRuO_3$ grown on 
slightly-miscut ($\sim 2^\circ$) substrates 
of $\rm SrTiO_3$. The orthorhombic film grows
with its $[001]$ and
 $[\bar{1}10]$ axes in the film plane and its  magnetic moment 
($\sim 1.4 \mu_B$ per ruthenium at saturation) along the out-of-the-plane axis of uniaxial
magnetocrystalline anisotropy (see Figure 1) whose direction varies in the (001) plane from
approximately 
$[010]$ ($45^\circ$ out of the plane) at $T_c$ to about $30^\circ$ relative to the normal to the film at
low temperatures.\cite{jcm} The residual resistivities of our films are as low as $ 4.6 \rm \ \mu
\Omega  \ cm$ (corresponding to resistivity ratio
 between $T= \rm 300 \ K$ and $T= \rm 1.8 \ K$ of $\sim 45$)
and their thicknesses vary
 between $800 \ \rm \AA$ and $2000 \ \rm \AA$. 
The films were
 patterned for simultaneous resistivity measurements  along  the $[001]$  and
 $[\bar{1}10]$ directions (see Figure 1).

The magnetic domain structure of $\rm SrRuO_3$ was
extensively studied using transmission 
electron microscopy (TEM) in Lorentz mode (see inset in Figure 1)
after removing the  $\rm SrTiO_3$ substrate with a chemical etch.
The details of this study are reported elsewhere;\cite{ann} here we mention the main
results relevant to this report: 
(a) there is a {\it single} easy axis along which the
spontaneous magnetization lies at zero applied field (no indication for flux closure
domains with a different magnetic orientation); (b) the DW are parallel to the
in-plane projection of  the spontaneous magnetization ($[\bar{1}10]$); therefore, each
time the sample is cooled through $T_c$ the DW appear in the same direction; 
(c) at low temperatures the spacing of the DW  is $\sim 2000 \rm \AA$ and it
does not change up to few degrees below $T_c$; (d)  once the DW are annihilated (at
temperatures lower than few degrees below $T_c$) by applying sufficiently-large magnetic field,
they do not renucleate when the field is set back to zero; 
only when a sufficiently-high negative field
 is applied,  new (less dense) domain walls are
nucleated in the process of magnetization reversal.

The effect of the DW on the resistivity was measured using two methods.  In the
first method the sample is cooled in zero field  from above $T_c$ to a temperature below
$T_c$. Then  the resistivity is measured during a hysteresis loop where the maximum field is high
enough to annihilate all domain walls. Figure 2 shows a hysteresis loop taken at $5
\rm \ K$.  We note that the initial zero-field resistivity (marked by full circles)  is
higher than the zero-field resistivity after the sample's magnetization was saturated.
  Based on our TEM measurements we know that the
initial resistivity here is measured in the presence of DW, while no DW are present in
the following zero-field resistivity measurements. 
Consequently, we attribute the difference between the two zero-field resistivities   to DWR.  In
the second  method the sample is cooled in zero field  from above $T_c$ to
$1.8 \rm \ K$ and the resistivity ($\rho^{ZFC}$) is measured as a function of temperature.
Afterwards, the sample is cooled from above $T_c$ in a field of $2 \rm \ T$ (to prevent
the formation of domain structure) down to
$1.8 \rm \ K$ where the the field is set to zero and the resistivity  ($\rho^{FC}$) is measured
as a function of temperature. Since $\rho^{ZFC}$ is measured with DW and $\rho^{FC}$ is
measured without them, we attribute the difference between the two to DWR.
 Figure 3 shows DWR as measured in the two methods as a function of
temperature for the two different current directions. However, before discussing these
plots (which are the main result of this paper), we address a crucial issue. Being
aware of contentions that previous reports on positive DWR originated  from anisotropic
magnetoresistance (AMR) effects, we find it vital to eliminate this possibility in our case.

The AMR effect is a dependence of the resistivity in a magnetic metal on the angle between 
the current and the magnetic moment which results from   
spin-orbit coupling. If in the domain state of a magnetic metal the
magnetization is along more than one axis, then saturating the magnetization of
the sample must induce in some parts of the sample changes in the angle
between the current and the magnetic moment; hence, an AMR effect. 
The existence of uniaxial anisotropy is not sufficient to ensure
that the magnetization points in the same direction  in all the domains.
It was pointed out\cite{exp2} that
when $Q=K/2\pi M_s^2 \ll 1$ (here $K$ is the anisotropy energy and $M_s$ is the
saturated magnetization), flux closure domains in which magnetization points in 
different directions are created near the surface of the sample; consequently, an AMR effect
contributes to the change of resistivity when the sample is saturated. While Fe and Co are in the
small $Q$ limit, $\rm SrRuO_3$ is in the  $Q\gg 1$ limit ($Q_{\rm SrRuO_3}>10$). Therefore,  it is not
surprising that contrary to Co and Fe, closure domains were not observed by TEM in $\rm
SrRuO_3$. This means that when we measure  the resistivity with or without DW the bulk
magnetization is along the same axis. This ensures not only the lack of any AMR effect but also any
effect related to changes  in the direction of the self field which can give rise to changes in the
regular Lorentz
 magnetoresistance.
 
After eliminating the possibility that  an intrinsic AMR effect contributes to our
measured DWR, we want to exclude the possibility of a 'dirt' effect; namely, that
small inclusions  of $\rm SrRuO_3$ with different magnetic anisotropy are causing the effect.
This is excluded not only by the low residual resistivity, and the consistent values among different
samples of the DWR in both current directions, but also by a direct test of the 
magnetic anisotropy of the regions responsible for the observed drop in resistivity
in the hysteresis loops. We  identify the
DWR in a finite field as the difference between the resistivity  on
the first branch of the loop (the initial increase of the field from zero to above the saturating field),
where  DW are partially annihilated, and the resistivity  on the second branch of the loop
(where the field is decreased from above the saturating field to zero), where no DW are present;
and we look at the dependence of the finite-field DWR on the angle  between the applied magnetic
field and the film. If the observed effect is a 'dirt' effect, we do not expect a preferred direction
(except for geometric considerations); on the other hand, if our interpretation
 is correct, we expect that the change in resistivity attributed to  the process of
annihilating the domain walls will depend on the component of the field parallel to the known
direction of the uniaxial anisotropy.   Figure 4 shows finite-field DWR for 
perpendicular current at some of the angles at which we measured. 
The inset shows that the
angular dependence  clearly supports our scenario. Similar results were also obtained for the
parallel current. Based on the above, we argue that indeed we measure DWR and not  
 a 'dirt' or a bulk effect. 

We go back now to Figure 3 which 
 shows  DWR as a function of temperature 
 for currents parallel
($\rho^\parallel_{DW}$) and perpendicular ($\rho^\perp_{DW}$) to the DW.
To the best of our knowledge, this is the most detailed temperature dependent measurement
of domain wall resistivity ever reported for either current direction.
The results presented in Figure 3 were obtained by measuring the sample with the lowest
residual resistivity, and
while results slightly vary among samples, we find the following characteristics in all our
measured samples. At low temperatures, 
$\rho^\perp_{DW}$  is always larger than $\rho^\parallel_{DW}$ and their
ratio is $\sim 2$ in the highest-quality samples. The magnitude of the DWR resistivity of various
samples in the zero temperature limit is very similar despite variations  of more than a factor of 2 in
the value of the residual resistivity. The specific features of the temperature dependence of
$\rho^\perp_{DW}$ and
$\rho^\parallel_{DW}$  are preserved. Up to $15 \ \rm K$, $\rho^\perp_{DW}$ is 
flat; between  $15 \ \rm K$ and $100 \ \rm K$ there is a sharp decrease in $\rho^\perp_{DW}$; 
between $100 \ \rm K$ and $120 \ \rm K$, $\rho^\perp_{DW}$ slightly increases; and between
$120 \ \rm K$ and $T_c$ $\rho^\perp_{DW}$ decreases to zero.
$\rho^\parallel_{DW}$  has very different behavior
except for the sharp decrease above $120 \ \rm K$ which is correlated with the sharp
decrease in the spontaneous magnetization.
 $\rho^\parallel_{DW}$  is
quite flat up to $120 \ \rm K$ with a shallow minimum around $60 \ \rm K$.
In the zero temperature limit we find that $\rho^\perp_{DW} \sim 0.48 \ \mu \Omega \ \rm cm$
and $\rho^\parallel_{DW} \sim 0.24 \ \mu \Omega \ \rm cm$. The width of the domain wall
is  $\sim 10 \rm \ \AA$ while the spacing between the domain walls is 
 $\sim 2000 \rm \ \AA$. Therefore, in terms of bulk resistivity, the resistivity for the
perpendicular current within the domain wall is   $\sim 100 \ \mu \Omega \
\rm cm$ and the interface resistance is $ \sim 10^{-15} \  \Omega
\ \rm m^2$. This value is more than three orders of magnitude larger than the $negative$ interface
resistance  of $ 6\pm 2 \times 10^{-19} \  \Omega
\ \rm m^2$ reported for cobalt.\cite{exp2} We believe that the huge difference in the magnitude of
the DWR  is  related to  the width of the
DW in cobalt estimated as 15 times larger than in $\rm SrRuO_3$.

Models proposed for DWR\cite{the1,the2,the3,the4,the5,gor} 
 have 
concentrated on the limit where the width of the DW is much larger than 
the Fermi wavelength. This limit, however, is not applicable here and therefore
we cannot compare our results to these models. Instead, we compare
our data to results obtained for magnetic multilayers. 
The configuration in which $\rho^\perp_{DW}$ is measured 
 is similar to the so-called CPP (current
perpendicular to plane) geometry in GMR structures (only that in our
case we study solely the effect of the magnetic interface without having to consider issues
such as surface roughness and matching between different materials). 
Therefore, mechanisms considered for the GMR structures may also be responsible for the
observed DWR.  
Such a mechanism is the potential step scattering, previously studied
for layered magnetic structures
by Barnas and Fert.
\cite{bf} The two found that an interface between a magnetic and a nonmagnetic metal, or
between different  magnetic domains within a magnetic metal  acts like
a potential step whose height is related to the exchange splitting. 
 While there is no closed form
equation for the interface resistance, the numerical solution of Barnas and Fert indicates that  for 
commonly used materials in magnetic
multilayers (e.g., Co or Cu) the  interface resistance is on the order of  $ \sim 10^{-15} \  \Omega
\ \rm m^2$.
Since DW in our case are of atomic width we 
 believe that it is possible to treat them as potential steps. The exchange splitting and the Fermi
energy in  $\rm SrRuO_3$ are $0.65 \rm \ eV$ and $2 \rm \ eV$, respectively.\cite{alsin}
Therefore,  we can expect an  interface resistance  on the order of  $ \sim 10^{-15} \  \Omega
\ \rm m^2$, as observed.  

Another potential source for DWR also considered for GMR structures is  spin
accumulation.\cite{fert2} When a polarized current crosses
an interface there is spin accumulation near the
interface that induces a potential barrier which results in excess resistivity.  
The spin accumulation (and its related resistivity) is strongly affected by the spin flip length
$l_{sf}$; namely, the length that a quasiparticle travels without flipping its spin. This length
is strongly affected by magnetic scattering; therefore, the decrease in  $\rho^\perp_{DW}(T)$
above $T=15 \ \rm K$ may be related to sharp decrease in the spin-accumulation resistivity
induced by the shortening of  $l_{sf}$ due to magnetic scattering. 

Contrary to $\rho^\perp_{DW}(T)$ which exhibits complex temperature dependence,
 $\rho^\parallel_{DW}(T)$ is almost flat  up to $120 \ \rm K$  despite big changes  particularly in
the resistivity but also in the magnetization of $\rm SrRuO_3$. 
We have no model for this  behavior; although, it is interesting to note
that we would have  obtained temperature-independent resistivity if we excluded the volume of
the DW including a distance from the DW proportional to the charge carrier mean free path. 

In conclusion, special properties of domain walls in $\rm SrRuO_3$ enable
clear observation of large $positive$ DW resistivity. 
Our main result here is the detailed temperature dependence of $\rho^\perp_{DW}$
and $\rho^\parallel_{DW}$ which requires further theoretical consideration. This, we hope,  will
yield deeper understanding not only of DWR but also of transport mechanisms in magnetic
multilayers.  Further experiments in which the mean free path will be changed by electron
irradiation, and the magnetization  will be changed by doping, are planned for  quantitative
identification  of the different contributions to DWR. 

We thank A. Fert and J. S. Dodge for useful discussions.
This research was supported by THE ISRAEL SCIENCE FOUNDATION  founded
by the Israel Academy of Sciences and Humanities and by
Grant No. 97-00428/1 from the United States-Israel Binational
Science Foundation (BSF), Jerusalem, Israel.

\begin{figure}
\caption{The measurement configuration: films are patterned for simultaneous measurement
of resistivity with parallel and  perpendicular domain walls (DW)
where the current is along the  $[\bar{1}10]$ and $[001]$  axes, respectively.  The DW are
 parallel to the
in-plane projection of the magnetization  vector $\bf M$.  Inset:  Image
of DW in $\rm SrRuO_3$ with Lorentz mode TEM. Bright and dark lines image
DW at which the electron beam  diverges or converges, respectively.
Background features are related to buckling of the free standing film and are not related to
magnetic variations.  }
\end{figure}

\begin{figure}
\caption{Hysteresis loops of resistivity $vs$ applied field for current parallel and perpendicular
to the domain walls at $T=5 \rm \ K$. At the starting point with $H=0$ (marked by full circle)
the sample is in its domain structure. Increasing the field annihilates the DW
and when the field is set back to zero
the magnetization of the sample  remains saturated. We identify
the difference between the initial zero-field resistivity and the subsequent zero-field resistivities as
the domain wall resistivity.  }
\end{figure}

\begin{figure}
\caption{Domain wall resistivity $vs$ temperature for
current parallel ($\rho^\parallel_{DW}(T)$) and perpendicular ($\rho^\perp_{DW}(T)$)
to the domain walls (along the $[\bar{1}10]$ and $[001]$ axes, respectively). Close symbols are
obtained by hysteresis loops (see Figure 2) and open symbols are obtained by difference between
$\rho^{ZFC}$  and $\rho^{FC}$. Inset: Temperature dependence of $\rho^{FC}$ 
for current along the $[\bar{1}10]$ (upper curve) and $[001]$ (lower curve)
axes and of the magnetization ($\bf M$). }
\end{figure}
  
\begin{figure}
\caption{Finite-field domain wall resistivity (see text) as a function of field at $T=10 \rm \ K$  when
the field is applied at different angles to the normal to the film. The inset shows 
the field needed to reduce the domain wall resistivity to $30\ \%$ of its full value
as a function of the angle between the applied field and the normal to the film.
The dashed line is a fit with $H=H_0 /|cos(\theta - 30)|$ which is the 
expected behavior if only the component of the field parallel to the uniaxial anisotropy
axis (whose orientation is at 30 degrees relative to to normal
to the film at that temperature) is relevant. 
 }
\end{figure}


\begin{thebibliography}{9999}

\bibitem{the1}  G. G. Cabrera and L. M. Falicov, Phys. Stat. Sol. {\bf 61}, 539 (1974);
{\it ibid}  {\bf 62}, 217 (1974).
\bibitem{the2} L. Berger, J. Appl. Phys. {\bf 49}, 2156 (1978).
\bibitem{the3} J. F. Gregg {\it et al}., Phys. Rev. Lett. {\bf 77}, 1580 (1996).
\bibitem{the4} G. Tatara and H. Fukuyama, Phys. Rev. Lett. {\bf 78}, 3773 (1997).
\bibitem{the5} P. M. Levy and S. Zhang, Phys. Rev. Lett. {\bf 79}, 5110 (1997).
\bibitem{gor} R. P. van Gorkom, A. Brataas, and G. E. W. Bauer, 
Phys. Rev. Lett. {\bf 83}, 4401
(1999).
\bibitem{exp1} G. R. Taylor, A. Asin, and R. V. Coleman, Phys. Rev. {\bf 165}, 621 (1968).
\bibitem{exp2} A. D. Kent {\it et al}., J. Appl. Phys. {\bf 85}, 5243 (1999);
U. Rudiger {\it et al}., Phys. Rev. Lett. {\bf 80}, 5639 (1998).
\bibitem{exp3} T. Taniyama {\it et al}., Phys. Rev. Lett. {\bf 82}, 2780 (1999).
\bibitem{loss} An expression for the domain wall width is  
$(C/2K_1)^{1/2}$. Here $K_1$
is the anisotropy constant which is  $\sim 10^7 \ \rm erg/cm^3$ (corresponding
to anisotropy field of $10 \rm \ T$) and
$C=2JS^2/a$ where $J$ is the exchange energy, $S$ is the spin and $a$ is the
distance
between the spins. The exchange $J$ is given by $k_BT_c/J=(5/96)(z-1)[11S(S+1)-1]$
where $z$ is the number of nearest neighbors. Plugging in $T_c=150 \rm \ K$, $z=6$ and 
$a=3.93 \ \rm  \AA$ we find a domain wall width on the order of $10 \rm \ \AA$.
\bibitem{jcm}  L. Klein {\it et al}.,
J. Phys. Condens. Matter  {\bf 8}, 10111 (1996) 
\bibitem{ann} A. F. Marshall {\it et al}., J. Appl. Phys. {\bf 85}, 4131 (1999).
\bibitem{bf} J. Barnas and A. Fert, Phys. Rev. B {\bf 49}, 12835 (1994).
\bibitem{alsin} P. B. Allen {\it et al}., Phys. Rev. {\bf B53}, 4393 (1996); 
D. J. Singh, J. Appl. Phys. {\bf 79}, 4818 (1996).
\bibitem{fert2} T. Valet and A. Fert, Phys. Rev. B {\bf 48}, 7099 (1993).

\end{thebibliography}
\end{document}